\documentclass[twocolumn,jgrga]{AGUTeX}
\usepackage{url}                                                

\usepackage[dvips]{graphicx}
\authorrunninghead{BISOI ET AL.}

\titlerunninghead{A 44 Hour Southward IMF-B$_{z}$}

\authoraddr{Corresponding author:Key Laboratory of Solar Activity, National Astronomical Observatories, 
Chinese Academy of Sciences, Beijing 100012, China. 
(e-mail: susanta@nao.cas.cn)}
\def\nat{{\it Nature}}
\def\jgr{{\it Journal of Geophysical Research (Space Physics)}}
\def\grl{{\it Geophys. Res. Lett.}}

\def\apj{{\it Astrophys. J.}}
\def\aa{{\it Astron. \& Astrophys. Lett.}}
\def\aap{{\it Astron. \& Astrophys.}}
\def\apss{{\it Astrophys. Space Sci.}}

\def\ssr{{\it Space Sci. Rev.}}
\def\solphys{{\it Sol. Phys.}}

\def\planss{{\it Planet. Space Sci.}}

\begin{document}

%
%

\title{A Prolonged Southward IMF-B$_{z}$ Event of May 02--04, 1998: 
Solar, Interplanetary Causes and Geomagnetic Consequences}

%
%


\authors{Susanta Kumar Bisoi,\altaffilmark{1}
D. Chakrabarty,\altaffilmark{2} P. Janardhan,\altaffilmark{3} 
R.G. Rastogi,\altaffilmark{4} A. Yoshikawa\altaffilmark{5}, 
K. Fujiki\altaffilmark{6}, M. Tokumaru\altaffilmark{6}, and 
Y. Yan\altaffilmark{1}}

\altaffiltext{1}{Key Laboratory of Solar Activity, National Astronomical 
Observatories, Chinese Academy of Sciences, Beijing 100012, China.}
\altaffiltext{2}{Space \& Atmospheric Sciences Division, Physical 
Research Laboratory, Ahmedabad 380009, India.}
\altaffiltext{3}{Astronomy \& Astrophysics Division, Physical Research 
Laboratory, Ahmedabad 380009, India.}
\altaffiltext{4}{Physical Research Laboratory, Ahmedabad 380009, 
India.}
\altaffiltext{5}{Department of Earth and Planetary Sciences,
International Center for Space Weather Science and Education, 
Kyushu University, 6-10-1, Hakozaki, Fukuoka 812-8581, Japan}
\altaffiltext{6}{ Institute for Space-Earth Environmental Research, 
Nagoya University, Nagoya 464-8601, Japan.}
%
%


\begin{abstract}
A detailed investigation was carried out to understand a 
prolonged ($\sim$44 hours) weakly southward interplanetary 
magnetic field (IMF-B$_{z}$) condition during May 02--04, 1998.  
{\it{In-situ}} observations, during the period, showed the 
passage of an expanding magnetic cloud embedded in an interplanetary 
coronal mass ejection (ICME), followed up by a shock and an 
interplanetary discontinuity driven by another ICME.  It is the 
arrival of the ICMEs and the upfront shocks that cause the prolonged 
southward IMF-B$_{z}$ condition.  The magnetic configuration of 
the source regions of the IMF associated with the ICME interval 
were also examined, which showed open magnetic field structures, 
emanating from a small active region on the north of the heliospheric 
current sheet (HCS).  The structures remained constantly to the 
north of the HCS, both on April 29 and May 01, suggesting no change 
in their polarity. The draping of these outward directed radial 
field lines around the propagating CMEs in the shocked plasma 
explained the observed polarity changes of the IMF-B$_{z}$ at 1 AU. 
In addition, multiple enhancements were also detected in the geomagnetic 
field variations, which showed a distinct one-to-one correspondence 
with the density pulses observed at 1 AU, during 0700--1700 UT on 
May 03.  The spectral analysis of both the variations showed the same 
discrete frequencies of 0.48, 0.65 and 0.75 mHz, demonstrating that 
the solar wind density enhancements can cause detectable global geomagnetic 
disturbances.  The observations, thus, provide a deeper insight into 
the possible causes and geomagnetic consequences of a prolonged weakly 
southward IMF-B$_{z}$ condition.
\end{abstract}

%
%

%

\begin{article}

%
%
%
\section{Introduction}
\label{S-Intro}
Inspite of significant advances in both instrumental capabilities and 
theoretical simulations, our understanding of various aspects of solar 
wind-magnetosphere-ionosphere coupling is still far from comprehensive.  
Nevertheless, over the years, a consensus has been reached on the 
potential impact of the north-south component of the interplanetary 
magnetic field (IMF-B$_{z}$), which controls the amount of solar 
wind energy transferred into the magnetosphere through magnetic 
reconnection of the IMF-B$_{z}$ and the terrestrial magnetic 
field (e.g., \cite{Dun61}).  The coupling is strongest when the 
IMF-B$_{z}$ is southward.  The solar plasma moving with a velocity, 
V${_{sw}}$ and having a frozen-in southward IMF-B$_{z}$ generates 
a dawn-to-dusk electric field, E${_{y}}$ = (-V${_{sw}}$ $\times$ Bz) 
\citep{TMe72,GoJ94,EcG08}. The electric field is primary responsible 
for causing magnetic storms.  It has been shown that it is the 
extraordinary and long duration southward IMF-B$_{z}$ events (magnitudes 
greater than -10 nT and lasting for over 3 hours), rather than high 
solar wind velocity, which plays a crucial role in triggering magnetic 
storms \citep{GTs87,TsL92}, and are therefore an important aspect 
for space weather studies.  

The causes of strong and long duration southward IMF-B$_{z}$ events identified 
to date are either of solar or interplanetary origins.  Events having 
solar origins are primarily associated with well-defined solar 
eruptive phenomena such as coronal mass ejection (CME) \citep{KBu82,LRL95}, 
solar flares \citep{TaT89}, filament or prominence eruptions \citep{TaT89}, 
and high-speed solar wind streams (HSS) from coronal holes (CHs) \citep{SHF76}. 
The major southward IMF-B$_{z}$ events responsible for severe, and great storms 
are all caused by magnetic clouds within CMEs \citep{GoB90,TsL92,EGT08}.  
A CME is a huge eruption of coronal plasma 
and is classically composed of three parts \citep{IHu86}. This includes 1) a leading 
edge 2) a dark cavity 3) a bright core.  The leading edge of the CME mainly comprises of 
coronal loops, and these have been detected at 1 AU \citep{TsA98}, while the bright 
core is due to the cold and dense filament material \citep{WHu87} which have also been 
detected at 1 AU.  The filament is geoeffective \citep{KoL14} while the loop 
is generally not \citep{TsE14}.  The dark region of the CME is known as magnetic cloud 
\citep{BuS81,KBu82} and can be detected at 1 AU in {\it{in-situ}} observations during the 
passage of the CME.  They are usually characterized by a higher magnetic field strength 
than the ambient solar wind, a radial dimension of $\sim$0.25 AU at 1 AU, a smoothly 
rotating field direction, and low proton temperature \citep{TsS88}.  More often all 
three components of a CME are not detected at 1 AU, and so in the interplanetary space a 
CME is known as interplanetary coronal mass ejection (ICME).  ICMEs at 1 AU 
were identified by observations of counter-streaming electrons with energies 
$\ge$ 80 MeV. This suggests that the ICME (with MC) have a magnetic flux rope structure 
\citep{KBu82} with their ends anchored to the Sun.  The radial fields rooted to the 
Sun are the ``magnetic foot'' regions of the ICME or the MC \citep{GoH74}. 

Events having interplanetary origins include those with shock compressed sheath 
fields \citep{TsS88,EcG08}, driver gas magnetic fields 
\citep{TaA85,TsS88,TaT89,EcG08}, draped interplanetary magnetic fields 
\citep{GMc87,TsS88,EcG08}, unusual or kinky heliospheric current sheets \citep{TsR84}, 
and corotating interaction regions (CIRs) \citep{SWo76,RCo80,TsH95,TsG06}. 
A small number of southward IMF-B$_{z}$ events are however, caused by large 
amplitude variations of intermittent Alfvenic turbulence, and waves or 
discontinuities \citep{TsS88,TsL92}. In a study of the interplanetary 
causes of the extreme southward IMF-B$_{z}$ events responsible for major 
geomagnetic storms during 1978--1979, \cite{TsS88} reported that half of 
the southward IMF-B$_{z}$ events were caused by shock compressed sheath fields, 
while the other half were caused by driver gas fields or magnetic clouds (MC).  
In a similar study during solar cycle 23, \cite{EcG08} also reported that 
intense southward IMF-B$_{z}$ events were caused by MC which 
drove fast shocks causing 24\% of the storms, sheath fields also causing 24\% 
of the storms, and combined sheath and MC fields causing 16\% of the storms. 
MC are thus considered to be an important source of the southward 
IMF-B$_{z}$ events. In a recent study, \cite{ZMo14} reported a statistical 
study of solar sources of extreme, long duration, southward IMF-B$_{z}$ events 
(IMF-B$_{z}$ $\geq$ 10 nT and lasting for more than 6 hours).  A majority (53\%) 
of these events were related to MCs and 10\% were related to ICME without 
MC signatures.  For events associated with MC having shocks 
ahead, if there is already pre-existing southward fields upstream of the shock, 
then the shock arrival compresses the fields and intensifies them \citep{TsS88,TsL11}.  
Subsequently, as these southward fields approach the MC, the field lines drape 
around the plasma leading to strong southward fields \citep{GoJ94}, thereby, 
resulting in a long duration southward IMF-B$_{z}$ event.  It is to be noted 
though, that the draping of the field lines around the CME depends on the CME 
trajectory, the direction of the radial interplanetary magnetic field (IMF), and 
the position of the heliospheric current sheet separating the positive and negative 
field lines \citep{GMc87,McG89}.

Though many attempts have been made, it is not straightforward to determine 
the possible cause of prolonged southward IMF-B$_{z}$ events, and it is certainly 
more difficult when the southward IMF-B$_{z}$ is weak for prolonged intervals.
The objective of the present paper is to investigate the cause of an extremely long 
duration ($\sim$44 hrs), weakly southward and relatively stable IMF-B$_{z}$ that 
gave rise to  geomagnetic storm conditions during the period May 2--4, 1998.  To the 
best of our knowledge, this is for the first time that such an unusually prolonged 
weakly southward IMF-B$_{z}$ and its possible solar or interplanetary 
causes is being reported.  However, there are reports of extended periods of southward 
IMF-B$_{z}$ events, such as the weakly southward IMF-B$_{z}$ for $\sim$12 hrs on 
February, 13--14, 2001 \citep{WaA06} and for $\sim$33 hrs on July 15--16, 2012 
\citep{BaH14}.  \cite{TaS05} also reported a long southward IMF-B$_{z}$ 
event of $\sim$31 hours while studying magnetotail response of prolonged southward 
IMF-B$_{z}$ events lasting longer than 8 hours during November 1999 through April 
2002.  However, none of these papers addressed their origin. 
\begin{table}[ht]
\vspace{0.0cm}
\begin{tabular}{| l | c | l |}
    \hline
    {\bf{Indian Stations}}     & {\bf{Magnetic Latitude (ML)}} \\ \hline
    Trivandrum (TRD)    & 00.03$^\circ$N \\  \hline
    Pondicherry (PND)   & 03.07$^\circ$N \\ \hline
    Visakhapatnam (VSK) & 08.56$^\circ$N \\ \hline
    Alibag (ABG)        & 10.36$^\circ$N \\ \hline
    Nagpur (NGP)        & 12.33$^\circ$N \\ \hline 
    Shillong (SHL)      & 16.30$^\circ$N \\ \hline
    Gulmarg (GUL)       & 25.60$^\circ$N \\ \hline  
    \end{tabular}
\caption{The seven Indian stations are listed in the increasing order of their 
magnetic latitudes, located mainly at low latitudes.  The measurements from the 
stations were used to show the multiple geomagnetic field variations, during the 
period from 0700--1700 UT on May 03, 1998 as seen in the left panel of 
Fig.\ref{deltaH}.}
\label{T-1}
\end{table}

The geomagnetic response of such extreme and long duration southward IMF-B$_{z}$ events 
are generally a magnetic storm \citep{TMe72,GTs87,TsL92} or a magnetic substorm 
\citep{TMe72,WaA06}.  However, during weakly southward IMF-B$_{z}$ fields, solar wind 
``ram'' pressure changes ($P_{dyn}$) become important.  Variations in $P_{dyn}$ commonly 
cause sudden impulses (SI) (e.g.,\cite{WSu61,RGP94b,FrL01,ViP05,VPi09}), and are generally 
caused by arrival of interplanetary shocks at 1 AU \citep{SCo65}.  On some occasions, 
pairs of sudden impulses or SI pairs, separated by a few hours, have been identified at 
ground stations \citep{RaJ10}. Also, the IP shocks (or ram pressure changes) could trigger 
substorms due to precursor southward IMF conditioning \citep{ZTs01}.  So, in general 
it can be said that, global geomagnetic disturbances, caused by such abrupt changes in 
solar wind velocity, have been of interest and have been studied in great detail over the 
years.  Besides velocity the ram pressure changes depends on solar wind density, which plays 
a role in ring current intensification \citep{CLS94,JoF98,BoT98,SmT99,WCL03}. Even the 
ram pressure changes arising due to only high density variations could externally trigger 
isolated intense substorm events or supersubstorms (AE $\leq$ -2500 nT) during storm or 
non-storm conditions \citep{TsH15}.  However, studies on the impact of solar wind density 
variations on the geomagnetic environment are sparse.  

The present prolonged southward IMF-B$_{z}$ event, on May 02--04, 1998, resulted in a mild 
storm on May 02 as well as an intense storm on May 04 \citep{GPF05}.  In addition to these 
storms, we also found quasi-periodic geomagnetic field variations for a duration from 0700 
UT--1700 UT on May 03, 1998.  Corresponding to these quasiperiodic magnetic field 
variations, similar solar wind density variations were observed at 1 AU.  So in the present 
investigation, based on a case study, we showed that solar wind density variations can 
affect geomagnetic field variations even during a long weakly southward and relatively 
stable IMF-B$_{z}$ condition.

It has also revealed by a spectral analysis of solar wind density variations 
having periodic or quasi-periodic fluctuations that they exhibit discrete 
frequencies \citep{KSS02,KSp03,VKS09} or wavelengths \citep{VKS08}.  These are 
particular frequencies or wavelengths, where the spectral analysis often shows 
significant power.  Such frequencies have been found to be occurred at 
0.7, 1.4, 2.0, 2.7 mHz in the ultra-low frequency (ULF) range. 
\cite{KSS02} and \cite{KSp03} also reported such discrete 
frequencies in magnetospheric fields, and concluded that the  ULF fluctuations 
in solar wind density were the driver of magnetospheric field oscillations.  Later, 
\cite{KSp03} argued that the density fluctuations instead were actually periodic density 
structures (PDS) formed near the solar surface, which being frozen-in the solar 
plasma, convect with the ambient solar wind to 1 AU.  In a study of solar wind density 
fluctuations for 11 years during solar cycle 23, \citep{VKS08} reported inherent 
radial scale sizes of PDS occurring at wavelengths of L $\sim$ 73, 120, 136, 
and 180 Mm (for the slow wind) and L $\sim$ 187, 270, 400 Mm (for the fast wind). 
To find the sources of PDS, a series of papers, based on the study of alpha-to-proton 
solar wind abundance ratio  \citep{VSK09}, and the remote observations of PDS using imaging 
data from {\it{Solar Terrestrial Relations Observatory}} ({\it{STEREO}}) spacecraft 
\citep{ViS10,VVo15}, reported that they are formed in the variable coronal solar plasma 
around or below 2.5 R$_{\odot}$, which advect with the slow solar wind to 1 AU.  In the 
present paper, we also verified whether such discrete frequencies were present 
in both solar wind density and geomagnetic field variations during the period 
from 0700 UT--1700 UT on May 03, 1998, when we found a clear one-to-one correspondence 
between the two. 
%
\begin{table}[ht]
\vspace{0.0cm}
\begin{tabular}{| l | c | l |}
    \hline
        {\bf{210 MM Stations}}     & {\bf{Magnetic Latitude (ML)}} \\ \hline    
    Kotel'nyy (KTN)     & 69.64$^\circ$N \\ \hline
    Tixie (TIK)         & 65.67$^\circ$N \\ \hline
    Chokurdakh (CHD)    & 64.67$^\circ$N \\ \hline
    Zyryanka (ZYK)      & 59.62$^\circ$N \\ \hline
    Magadan (MGD)       & 53.56$^\circ$N \\ \hline 
    Moshiri (MSR)       & 37.61$^\circ$N \\ \hline
    Popov Island (PPI)  & 36.62$^\circ$N \\ \hline 
    Onagawa (ONW)       & 24.85$^\circ$N \\ \hline
    Kagoshima (KAG)     & 25.13$^\circ$N \\ \hline
    Yamakawa (YMK)      & 13.80$^\circ$N \\ \hline
    Lunping (LNP)       & 10.36$^\circ$N \\ \hline
    Guam (GAM)          & 04.57$^\circ$N \\ \hline 
    Muntinlupa (MUT)    & 03.58$^\circ$N \\ \hline
    Biak (BIK)          & 12.18$^\circ$S \\ \hline 
    Learmonth (LMT)     & 34.15$^\circ$S \\ \hline
    Canberra (CAN)      & 45.98$^\circ$S \\ \hline 
    Katanning (KAT)     & 46.03$^\circ$S \\ \hline  		
    \end{tabular}
\caption{The seventeen stations of 210 MM network and their magnetic latitudes are 
shown. The stations are located at latitudes starting from low though mid and on to high 
latitudes.  Thirteen of them are in the northern hemisphere, while four of them are in the 
southern hemisphere.  The measurements from the stations were used to show the multiple 
geomagnetic field variations, during the period from 0700--1700 UT on May 03, 1998, as 
seen in the right panel of Fig.\ref{deltaH}.}
\label{T-2}
\end{table}
%
\subsection{Tracking of Solar Wind Events at 1 AU to its Solar Source Region}
Using {\it{in-situ}} measurements, solar wind flows can be traced back, from 
the Earth's orbit (at 1 AU) to a solar source surface at 2.5 R$_{\odot}$, where 
R$_{\odot}$ is solar radius.  Further, using measurements of photospheric magnetic 
fields and potential field computations the flows can be mapped back to the Sun. 
Many examples exist of such studies, some of which are listed here.  In order to 
study quadrupole distortions of the heliospheric current sheet, from May 1976 to 
May 1977, \cite{BBH82,BHu84} successfully used the solar wind trace-back method 
in conjunction with a potential field model.  In early 1995, using data obtained from 
the Ulysses and Wind spacecraft, \cite{NeF98} traced back solar-wind structures
to the source surface at 2.5 R$_{\odot}$ and then mapped them back to the photosphere.
In an attempt to link outflows from an equatorial coronal hole and solar wind observed 
at 1 AU, \cite{MLD11} used space based data from the 
{\it{Advanced Composition Explorer}} ({\it{ACE}}) to trace the solar wind back to the 
source surface using potential field source  surface (PFSS) models.  To identify the 
source regions of gradual solar energetic particle (SEP) events at 1 AU, \cite{KoT13} 
first used solar wind speed measurements to map the Sun-L1 interplanetary magnetic 
field lines back to its source regions on the Sun at the time of SEP observations, 
and compared the characteristics of the SEP events to the identified sources of IMF. 

The velocity traceback technique though, is applicable only to steady state solar 
wind flows, however, it has also been successfully used even when the solar wind flow 
were highly non-radial (e.g. \cite{BaJ03,JaF05,JaF08,JDM08}) during the well known 
“solar wind disappearance event” of May 11, 1999. To identify the solar source locations 
of the disappearance event, the authors traced back solar wind flows to the source 
surface at 2.5 R$_{\odot}$, and using potential field computations \citep{HKo99} 
mapped them back to the photosphere. From their study of the disappearance event of May 
11, 1999, \cite{JaF05} reported the maximum trace back errors to be $\sim$30$^{\circ}$. 

In this paper, we also used the two-step trace back method to locate 
the footpoints or the source region of the IMF associated with the prolonged 
IMF-B$_{z}$ event.  In the first step, we traced back the field lines, using 
{\it{in-situ}} measurements of solar wind speed from the {\it{ACE}} 
spacecraft at 1 AU, to the source surface at 2.5 R$_{\odot}$.  Following this, 
in the second step, using potential field computations \citep{HKo99} to synoptic 
maps, we traced the field lines from the source surface back to the photosphere, 
and investigated the background magnetic field conditions on the Sun during our 
period of interest.

It is important to note that generally in such potential field models, the 
coronal magnetic field below the source surface is assumed to be quasi-stationary, while 
it is considered to be open beyond the source surface. Also, it is to be noted that the 
tracing of the source region and magnetic field lines, using a Carrington synoptic 
magnetogram has to be carefully treated at the two ends of a magnetogram. 
The magnetic field at this region changes faster than the normal level, which can lead 
to variations in the tracing process and the extrapolations of the magnetic field 
lines \citep{HuY14}.  However, for our May 2--4, 1998 event, the Carrington longitude 
was close to the center of the magnetogram, so the boundary effect would not be 
significant.
\section{Presentation of Data}
\label{S-data}
For IMF and solar wind observations at 1 AU, we have used data available in the 
public domain from \url{http://cdaweb.gsfc.nasa.gov/cgi-bin/eval2.cgi}.  For 
interplanetary plasma and IMF data such as solar wind proton velocity and the 
strength of IMF we have used data obtained from the {\it{ACE}} spacecraft 
\citep{StF98}.  Ground based measurements of solar wind velocities in the inner heliosphere, 
using radio observations of interplanetary scintillation at 327 MHz \citep{KKa90} 
were also utilized.  For studying the photospheric magnetic field distribution, synoptic 
magnetograms, made using data from the {\it{Michelson-Doppler-Interferometer}} ({\it{MDI}}: \cite{ScB95}) 
instrument onboard the {\it{Solar and Heliospheric Observatory}} ({\it{SOHO}}: \citep{DFP95}), 
were used. Measurements of ground magnetic deviations were obtained from  
both Indian stations and the 210$^{\circ}$ magnetic meridian (210 MM) network 
\citep{YCp01}. The seven Indian and seventeen 210 MM stations used have been shown in Tables  
\ref{T-1} and \ref{T-2}.  The Indian stations are mainly located at 
low-latitudes, while the 210 MM network of stations are located at latitudes 
starting from low though mid and on to high latitudes.  Of these seventeen stations, 
thirteen are located in the Northern hemisphere, while four are in the Southern hemisphere.
\subsection{Interplanetary Scintillation Observations}

Interplanetary Scintillation (IPS), typically observed at meter wavelengths, 
is a diffraction phenomenon in which coherent electromagnetic radiation from 
a distant radio source passes through the turbulent, refracting solar wind and 
suffers scattering, thereby resulting in random temporal variations of the signal 
intensity (scintillation) at the Earth.  A schematic illustration of the typical
IPS observing geometry can be seen in \cite{BiJ14b}.  IPS is an efficient, 
cost-effective method of studying the large scale structure of the solar wind at meter 
wavelengths, over a wide range of heliocentric distances, ranging from 
about 0.2 AU to 0.8 AU at 327 MHz \citep{HSW64,JAl93,ABJ95,JaB96,MoA00,BiJ14b}.  Intensity 
variations from point-like extragalactic radio sources, during IPS observations, 
actually provide remote-sensing information about rms electron density fluctuations, 
$\Delta{N_{rms}}$, along the lines-of-sight to the observed radio sources 
\citep{BiJ14b}.  In fact, IPS is so sensitive to changes in $\Delta{N_{rms}}$ that 
it has even been used to study the fine-scale structure in cometary ion tails during 
radio source occultations by cometary tail plasma \citep{JaA91,JaA92,IjA15}.  IPS data were 
obtained from the Institute for Space-Earth Environmental Research (ISEE), Nagoya University, 
Nagoya, Japan and using a tomographic technique, IPS observations from ISEE, Japan 
were used to produce synoptic velocity maps \citep{KoT98} which were then projected back to a fixed 
heliocentric distance, {\it{viz.}} the source surface, at 2.5 R$_{\odot}$.  Such 
synoptic velocity maps represent the large-scale structure of the solar wind in the inner 
heliosphere.
\section{A Prolonged Southward IMF-B$_{z}$ Event}
\label{S-IMF}
\begin{figure}[ht]
\vspace{13.3cm}
\includegraphics{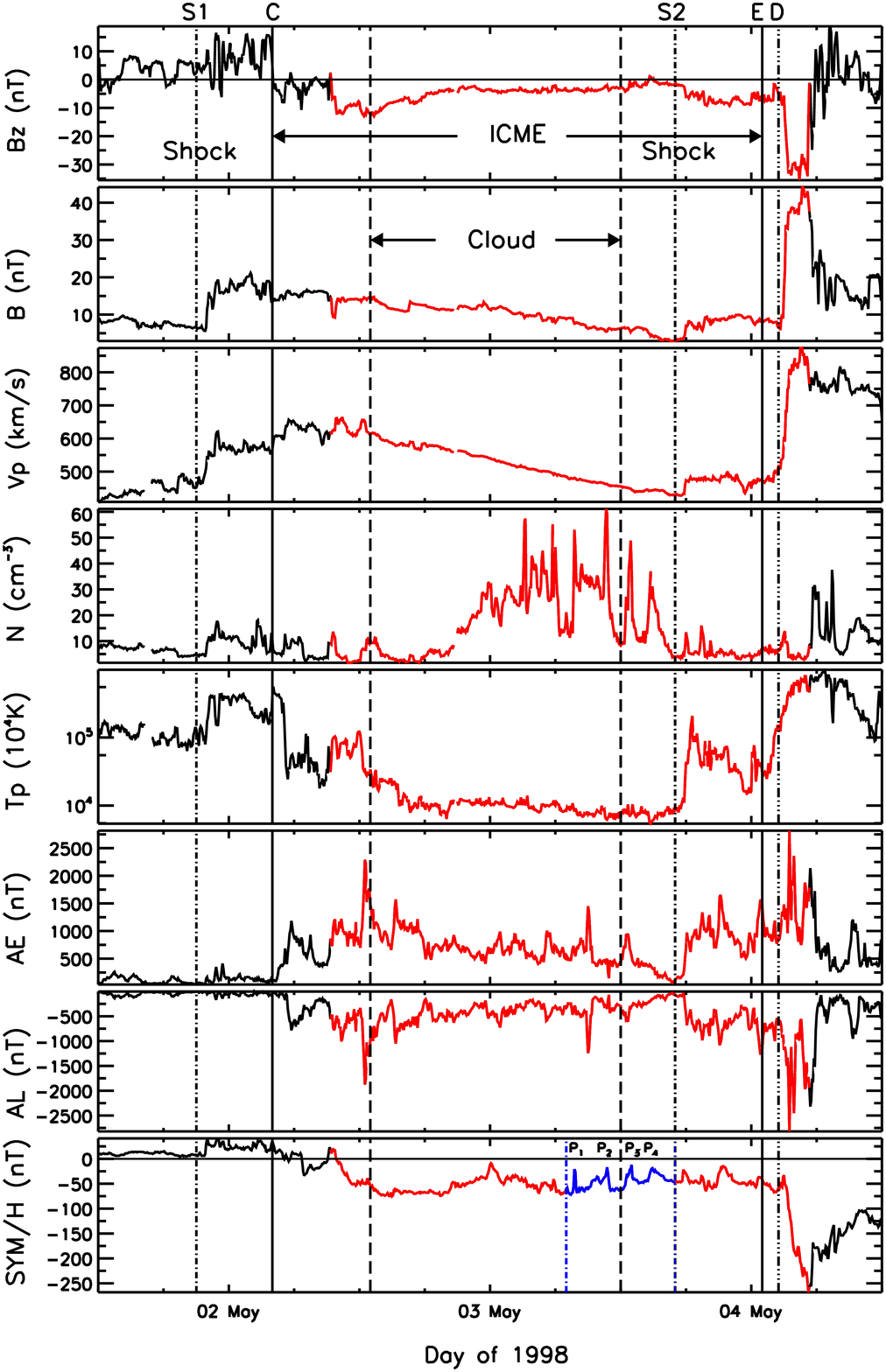}
\caption{From top to bottom, panels show spacecraft measurements of IMF-B$_{z}$, magnitude 
of magnetic field, proton velocity, density and temperature, and ground measurements of 
AE, AL and SYM-H indices, respectively, on day of 1.5\,--\,4.5 on May, 1998. The two vertical 
dash-dotted lines (S1 and S2) and the vertical long dash-dotted line (D), respectively, mark 
the arrival of shocks and discontinuity, while the two solid lines (C and E) mark the arrival 
and end of CME, respectively.  The vertical dashed lines indicates the start and end of 
magnetic cloud. An unusually prolonged southward IMF-B$_{z}$ for a period of $\sim$44 hours 
is shown by the red solid line.  The multiple enhancements observed in SYM-H index, during 
0700 UT--1700 UT on May 03, have been shown by the blue solid line, indicated between the blue 
vertical dash-dotted lines.}
\label{fig1}
\end{figure}

Fig.\ref{fig1} presents solar wind conditions at 1 AU (measurements of solar wind 
magnetic field and plasma) of an event on day of 1.5\,--\,4.5 on May, 1998, and the 
state of magnetospheric response associated with the event.  The topmost panel of 
Fig.\ref{fig1} shows the southward component of interplanetary magnetic field 
(B$_{z}$).  An unusually prolonged southward IMF-B$_{z}$ lasting for a period of 
$\sim$44 hours, from 0920 UT on May 02 to 0520 UT on May 04, is clearly evident as 
shown by the solid red line in the uppermost panel of Fig.\ref{fig1}.  To identify 
the cause of such a prolonged southward IMF-B$_{z}$ condition, we first studied 
the interplanetary (IP) disturbances associated with the event through solar 
wind parameters such as magnitude of magnetic field (B), proton velocity (V$_{p}$), 
proton density (N$_{p}$) and proton temperature (T$_{p}$) as shown from the top 
second to the fifth panel of Fig.\ref{fig1}, respectively. 
%

A forward shock was identified at 2123 UT on May 01 as indicated by a vertical dash-dotted line (S1).  
The arrival of the shock increased the proton density ($\sim$20 cm$^{-3}$), velocity (600 kms$^{-1}$), 
and temperature ($\sim$10$^{10}$K) as well as magnetic field intensity ($\sim$20 nT).  Prior to the shock, 
the magnitude of IMF was high ($\sim$10 nT), and the IMF-B$_{z}$ was $\sim$0 nT 6 hours before and 
$\sim$+7 nT just 30 min before. The solar wind plasma preconditions were as follows: B was $\sim$10 nT, 
V$_{p}$ was $\sim$ 300 kms$^{-1}$, and N$_{p}$ was $\sim$ 10 cm$^{-3}$.  Post shock, the IMF-B$_{z}$ showed 
rapid fluctuations, and the average IMF-B$_{z}$ was  northward having $\sim$+10 nT. However, the 
field actually turned first southward (B$_{z}$ was $\sim$-8 nT) from the northward direction (B$_{z}$ 
was $\sim$10 nT)  at 0300 UT on May 02 upon arrival of an ICME at 1 AU, marked by a vertical solid line 
(C).  The arrival of ICME at 1 AU has been characterized by jumps in V$_{p}$, N$_{p}$ and T$_{p}$, and a 
decrease in B.  However, until 0920 UT, the southward field was not stable, and showed a second southward 
turning only at 0920 UT on May 02.  Thereafter, the IMF-B$_{z}$ was constantly southward and relatively 
stable for the next 24 hours. 
%
\begin{figure}[ht]
\vspace{8.4cm}
\includegraphics{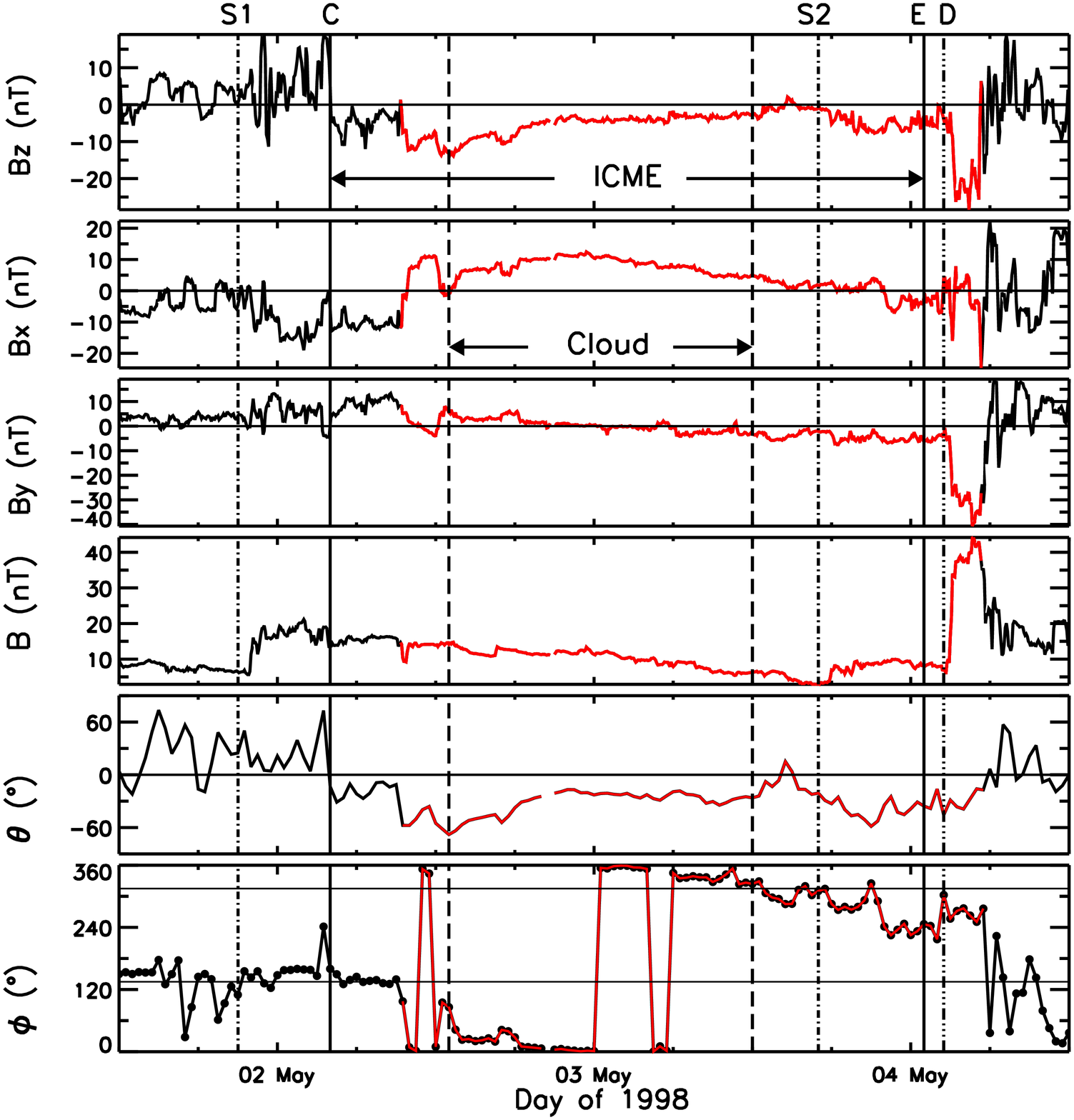}
\caption{{\it{In-situ}} measuements of magnitude and direction of the magnetic field.  
From top to bottom, panels show B$_{z}$, B$_{x}$, B$_{y}$, B, $\theta$, and $\phi$, 
respectively, on day of 1.5\,--\,4.5 on May, 1998.  The two vertical 
dash-dotted lines (S1 and S2) and the vertical long dash-dotted line (D), respectively, 
mark the arrival of shocks and discontinuity.  The two solid lines (C and E) 
mark the arrival and end of CME, respectively, while the vertical dashed lines 
indicate the start and end of magnetic cloud.  The steady state configuration of the 
IMF with $\phi$ = $135^{\circ}$ or $315^{\circ}$ is shown by the solid lines in the 
bottommost panel.}
\label{fig2}
\end{figure}	      

The solar wind plasma during the prolonged period of weakly southward IMF-B$_{z}$ condition showed 
constant high magnetic field intensity than the ambient field and continuous low proton temperature. 
These are typical signatures of the presence of a driver gas plasma or magnetic cloud (MC). The passage 
of the MC \citep{MaS02} is indicated by vertical dashed lines in Fig.\ref{fig2}.  \cite{MaS02} 
reported two prompt electron events as detected by {\it{ACE}} spacecraft during the passage of ICME, 
one when the CME leading edge arrives at 1 AU and the other when the front of MC passes at 1 AU.  
The monotonic decreases in both magnetic field and proton speed suggest that the ICME was expanding.  
Since the MC was traveling for around 24 hours with an approximately speed of 500 kms$^{-1}$, it 
suggests that the radial dimension of the MC was $\sim$0.28 AU, which is the typical radial width of 
a MC as mentioned earlier.  Also, a high density plasma parcel (30-60 cm$^{-3}$) was located between 
1800 UT on May 02 and 1700 UT on May 03, for about 24 hours.  The density parcel, in fact, showed 
periodic/quasi-periodic density variations.  The He$^{++}$/H$^{+}$ ratio (though not shown here) 
had also shown significantly increased values, as high as 80\% \citep{SkB99}. 

A second forward shock was identified at 1700 UT on May 03, indicated by a vertical dash-dotted 
line (S2).  The arrival of shock showed a small jumps in V$_{p}$, N$_{p}$, T$_{p}$, and B 
compressing the already pre-existing southward fields and intensifying them further.  Prior to the 
shock, the solar wind plasma preconditions were as follows: B was $\sim$0 nT, V$_{p}$ was $\sim$ 
300 km$s^{-1}$, and N$_{p}$ was $\sim$ 8-10 cm$^{-3}$. The IMF-B$_{z}$ was $\sim$0 nT just 2 hours before. 
The shock was propagating back of the first ICME, and was associated with a second ICME arriving later 
\citep{MaS02}.  The end of ICME interval (E), shown by a vertical solid line, was observed with a 
decrease in T$_{p}$.  Just prior to the end of ICME, when the southward field was slowly turning 
northward, a strong interplanetary discontinuity was identified, indicated by a vertical long 
dash-dotted line (D). Post-discontinuity, the solar wind showed high jumps in V$_{p}$, N$_{p}$, T$_{p}$, 
and B, however, these variations occurred over an extended period suggesting that it was not a shock.  
The discontinuity compressed the weak pre-existing southward fields and intensifying them further 
(B$_{z}$ decreases from -5 nT to -32 nT).  The variations in solar wind plasma composition and 
properties during May 02-04, 1998 have been very well studied and reported by other researchers (e.g., 
\cite{GlF99,SkB99,CFr99,CFr01,BaW02,MaS02,PoE03,GPF05}).
\subsection{Orientation of the Magnetic Fields}
\label{S-mag}
Fig.\ref{fig2} shows, from top to bottom, the magnetic field components 
(B$_{z}$, B$_{x}$, B$_{y}$) in Geocentric Solar Ecliptic (GSE) coordinates, the magnitude 
of magnetic field (B), and the direction of magnetic field ($\theta$ and $\phi$).  
In the GSE coordinate system, the X axis points from the Earth to the Sun, the Z axis is 
normal to the ecliptic, and the Y axis forms a right hand coordinate system. The angle 
$\theta$ is the angle of the magnetic field out of the ecliptic plane, while the angle 
$\phi$ is the angle that the magnetic field makes with the X-axis in the XY plane. Before 
the passage of the MC, the magnetic field showed rapid fluctuations, while it was very smooth 
during its passage, as indicated by the low variance in the magnitude of magnetic field.  
From Fig.\ref{fig2}, it is clear that the B$_{x}$ component remained positive during MC 
traversal, while the B$_{z}$ component was negative.  However, the B$_{y}$ component 
showed a change from positive to negative, indicating the usual directional changes 
noticed during the MC traversal.  The typical Parker's spiral configuration of the IMF has values 
of $\phi$ $=$ $135^{\circ}$ or $315^{\circ}$, which corresponds to the positive or the 
negative IMF, respectively, shown by the horizontal lines in the bottommost panel of 
Fig.\ref{fig2}.  The value $\theta$ $=$ $90^{\circ}$ usually denotes the position of the ecliptic 
north pole.  It is seen from Fig.\ref{fig2} that $\theta$ was negative during the whole 
event.  The IMF turned southward (negative $\theta$), upon the arrival of the ICME, and 
continued to be thereafter southward until May 04. The angle $\phi$ was around $135^{\circ}$ 
from 00 UT to 08 UT on May 02, indicating a steady state configuration with a positive IMF. 
The first crossing of the IMF was noticed just prior to the MC and on it's arrival, the IMF 
became non-steady for a while.  The second crossing of the IMF was seen early on May 03,  
from positive to negative, corresponding to the directional changes in the B$_{y}$ component. 
Thereafter, the angle $\phi$ was continued to be around $315^{\circ}$ until 1700 UT on 
May 03, indicating a steady negative IMF configuration, which again became non-steady 
upon the arrival of the shocks driven by the second ICME from 1700 UT on May 03 to early on 
May 04.
%
\begin{figure}[ht]
\vspace{12.6cm}
\includegraphics{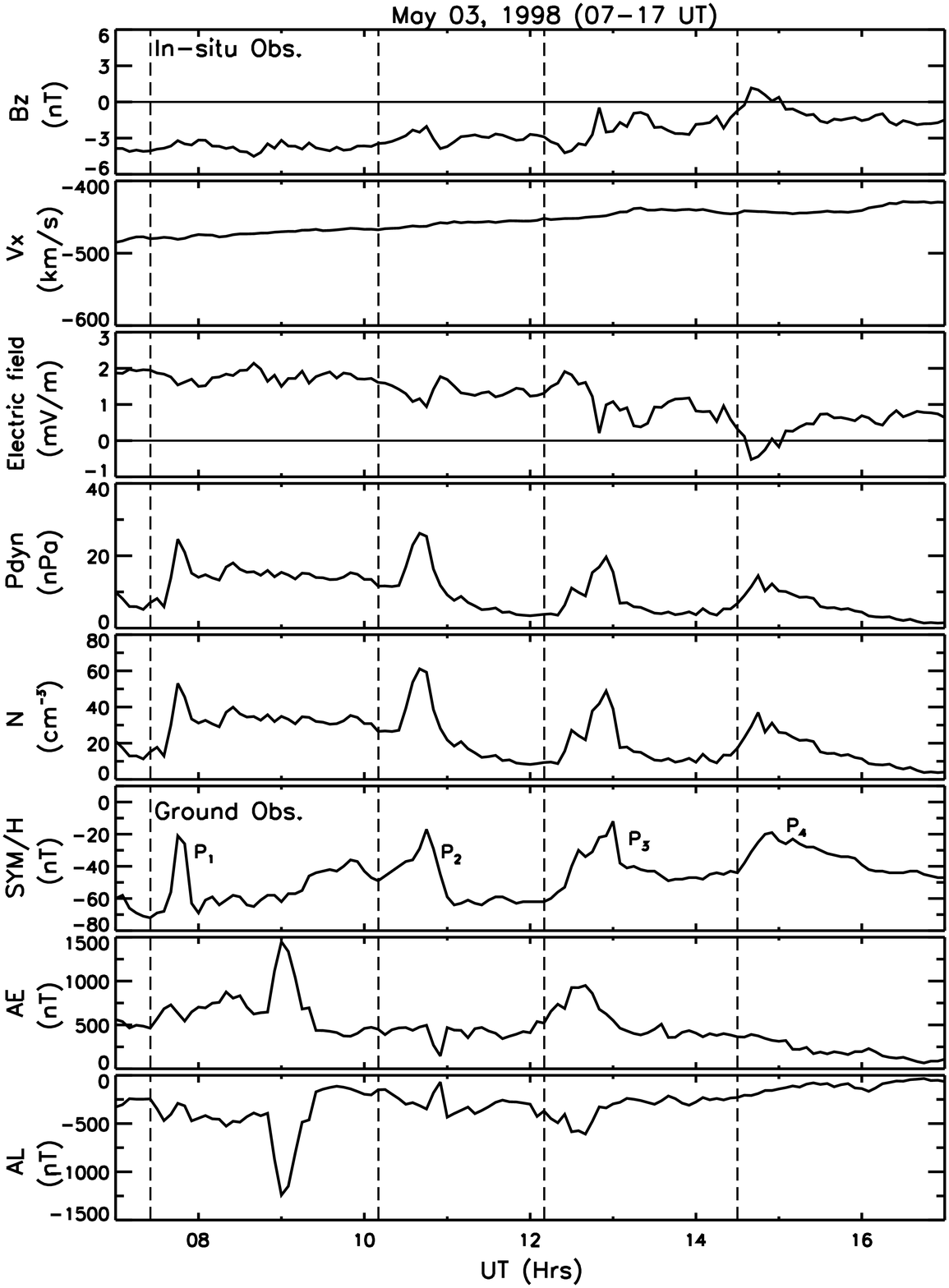}
\caption{The panels starting from the top show variations of the B$_{z}$, solar wind 
velocity, electric field, flow pressure, solar wind density, SYM-H, AE, and AL 
as a function of time in UT on May 03, 1998, respectively. The vertical dashed 
lines demarcate the occurrence of multiple enhancements in SYM-H when there are no 
corresponding significant variations in both the IMF-B$_{z}$ and solar wind velocity. 
The multiple peaks in SYM-H are numbered from P$_{1}$ through P$_{4}$, and corresponding 
to these pulses, multiple enhancements in solar  wind density and solar wind dynamic 
pressure have also been identified.} 
\label{fig3}
\end{figure}	      
%
%
\subsection{Geomagnetic Consequences}
\label{S-magres}
The magnetospheric response corresponding to the unusually prolonged southward 
IMF-B$_{z}$ condition on May 02--04 has been shown in the bottom three panels of 
Fig.\ref{fig1}, which show ground measurements of the auroral electrojet index (AE), 
the westward electrojet index (AL) and the symmetric ring current index (SYM-H), 
respectively.  On arrival of the first forward shock, S1, a sudden increase in SYM/H 
index ($\geq$ 25 nT), {\it{i.e.}}, a positive sudden impulse (SI), was observed.  
However, only a very small increase in AE and AL indices was noticed.  The arrival of 
the ICME, that caused the first southward turning of the IMF-B$_{z}$ field and the 
reverse shock, initiated a depression in SYM-H index early on May 02 at 0400 UT.  So 
the arrival of forward and reverse shocks caused, respectively, a positive and a 
negative SI, as shown in the study by \cite{RaJ10}.  The AE and AL indices, 
respectively, showed increases up to $\sim$1000 nT and $\sim$-800 nT indicating onset of a 
moderate substorm activity.  A moderate storm (SYM-H $\leq$ -80 nT) actually began on the 
second southward turning of the IMF-B$_{z}$ when the southward IMF-B$_{z}$ field intensified 
(B$_{z}$ $\leq$ -10 nT) further.  Correspondingly, AE and AL indices showed much higher values, 
indicating the progression of a substorm activity.  At about 1300 UT on May 02, their amplitudes 
had values greater than -1500 nT, indicating an intense substorm activity in progress.  Besides, 
the southward pre-conditioning, the solar wind plasma density was found to be higher 
($\sim$10-15 cm$^{-3}$), coincident with the substorm event.  The increased solar wind density 
could have externally triggered the intense substorm activity.

Thereafter, on weakening of the southward field, the storm entered into the recovery phase 
as indicated by the slow increase in SYM-H index.  The passage of a high density plasma 
parcel early on May 02 further caused a depression in SYM/H index.   In fact, multiple 
enhancements in SYM-H index were identified, shown by the blue solid line and demarcated by 
the vertical blue dash-dotted lines in the bottommost panel of Fig.\ref{fig1}, from 0700 
UT--1700 UT on May 03, during the steady weakly southward IMF-B$_{z}$ condition. 
The multiple peaks in SYM-H occurred at 0725 UT, 1010 UT, 1210 UT, and 1430 UT and were 
referred to in the rest of the paper as P${_1}$, P${_2}$, P${_3}$, and P${_4}$, respectively.  
During this steady weakly southward IMF-B$_{z}$ condition, we found fair amount of 
constant changes in both AE and AL indices indicating a fair substorm activity.  The arrival 
of the second forward shock again caused sharp increases in AE and AL indices that led to the 
onset of a moderate substorm activity.  Subsequently, the arrival of the IP discontinuity 
caused a strong intensification of the southward IMF fields, and resulted in an intense 
geomagnetic storm that began early on May 04.  Correspondingly, the SYM-H index showed a 
strong positive SI followed by a strong depression in its strength ($\leq$ -250 nT).  Also, 
we found substantial enhancements in AE index ($\geq$ 2500 nT) as well as in AL index 
($\leq$ -2500 nT) that led to a large substorm expansion.  

It is thus clear that during the prolonged southward IMF-B$_{z}$ condition of May 02--04, a 
constant geomagnetic activity with the values of SYM-H $\leq$ -50 nT and a fairly constant 
substorm activity with the values of AE $\geq$ 500 nT was observed prior to the onset of the 
great storm on May 04. The storms during May 02--04 have been very well studied 
\citep{CFr99,CFr01,BaW02,PoE03,GPF05}.  However, the multiple enhancements noticed in SYM-H, 
during the prolonged quasi-stable southward IMF-B$_{z}$ condition on May 03, are unique and 
are being reported for the first time.  We discuss them in detail in the following section.
%
\protect
\begin{figure*}[ht]
\vspace{9.5cm}
\centering
\includegraphics{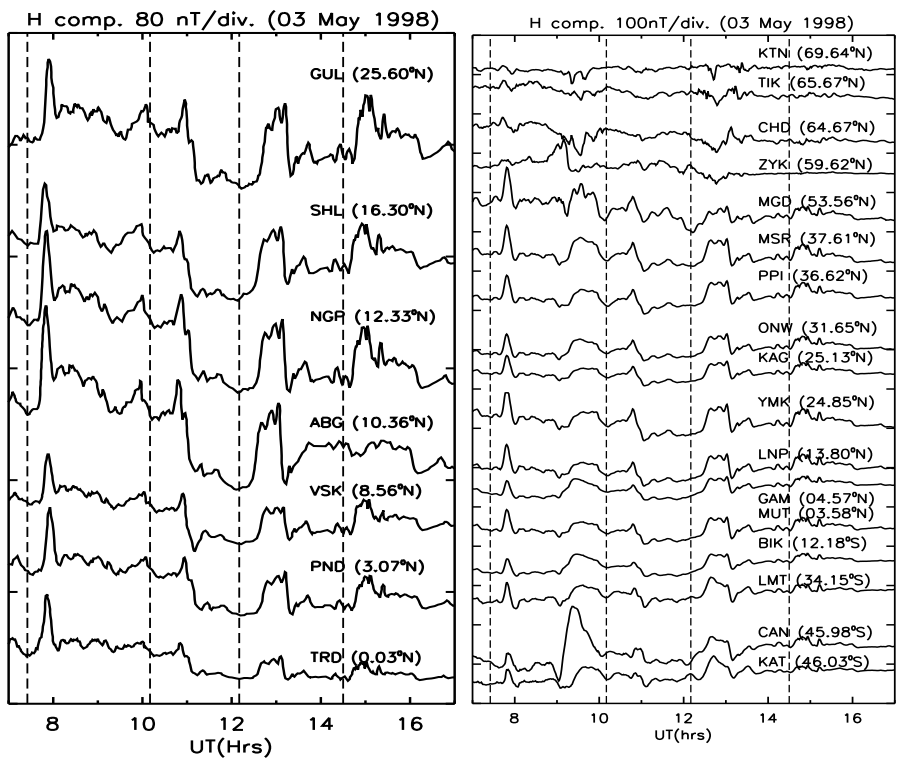}
\caption{Variations of the horizontal component (H-comp) of the Earth's magnetic 
field for Indian stations (left panel) and 210 MM stations (right panel) between 
0700 UT and 1700 UT on May 03, 1998.  Shown on the right of each curve is the 
geomagnetic latitude of each station.  The vertical dashed lines have been marked 
at the time of onset of pulses P$_{1}$, P$_{2}$, P$_{3}$, and P$_{4}$, respectively, 
in the SYM-H field.}
\label{deltaH}
\end{figure*}
\section{Multiple Pulses in the Geomagnetic Field Variations}
\label{S-multi-pulse}
As mentioned earlier, as a geomagnetic response to the quasi-stable southward 
IMF-B$_{z}$ condition during May 03, 1998, we identified the occurrence of 
multiple enhancements in the Earth's magnetic field as seen by the multiple 
pulses in SYM-H index during 0700 UT--1700 UT on May 03.  Starting from the 
top and going down, the panels of Figure \ref{fig3} show respectively, 
the IMF-B$_{z}$, the component of solar wind velocity in the X-direction (V$_{X}$), 
the Y-component of the interplanetary electric field (IEF$_{y}$), the solar wind 
dynamic pressure (P$_{dyn}$), the solar wind proton density (N), the SYM-H index, 
the AE index, and the AL index for the period 0700 UT--1700 UT on May 03.  
The multiple pulses are numbered from P$_{1}$ through P$_{4}$, and have also 
been shown in the lowermost panel of Fig.\ref{fig1}. 

Corresponding to the multiple peaks in SYM-H, we identified multiple enhancements 
in solar wind dynamic pressure and solar wind density, indicated by the dashed vertical 
lines in Fig.\ref{fig3}.  These density variations are specifically of interest, 
as there were no significant variations observed in the IMF-B$_{z}$, solar 
wind velocity, and the IEF$_{y}$ except for the corresponding deviations in the 
SYM-H.  We also didn't see much variation in both AE and AL indices corresponding to 
the peaks in the SYM-H field. It is important to note that the surface magnetic 
deviations in the SYM-H have a distinct one-to-one correspondence both in time and 
profile with that of the solar wind density pulses. Since the SYM-H field represents 
the mean variation of the H field at all mid-latitude ground stations, the observed 
deviations showed that the event is a global one.  Further, we have taken care to verify 
that the SYM-H pulses are global by examining records of the change in the horizontal 
component of geomagnetic field ($\Delta$H) from ground-based magnetograms obtained 
from Indian magnetic observatories as well as from the 210 MM magnetic stations.
%
%
\protect
\begin{figure}[ht]
\vspace{7.6cm}
\centering
\includegraphics{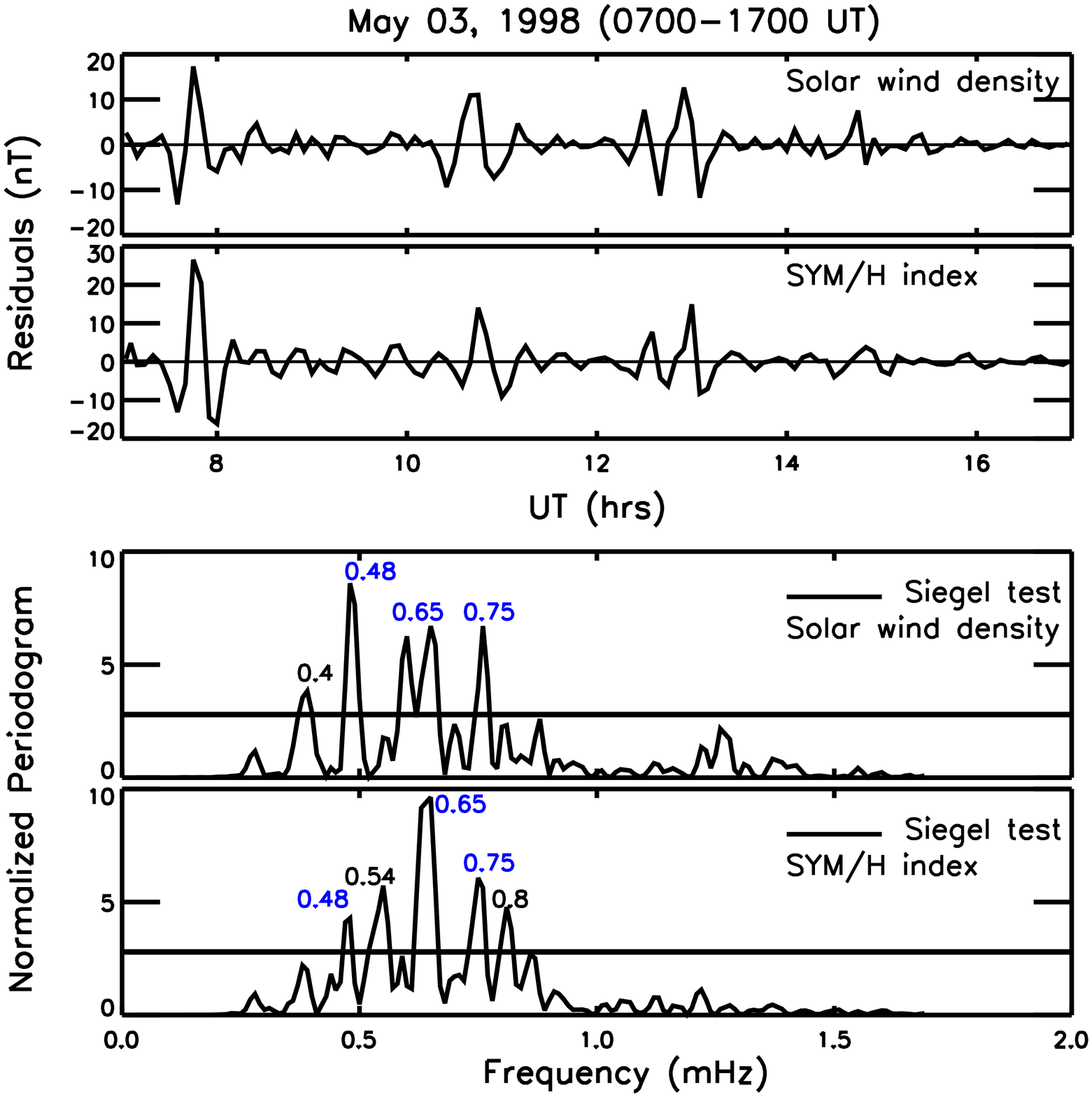}
\caption{The top two panels show variation of the residuals of solar wind density 
and SYM-H field during the period from 0700--1700 UT on May 03, while the 
Lomb-Scargle normalized periodograms of them are shown in the bottom two panels. 
The black horizontal lines, shown in the bottom two panels, are drawn at a 
critical level determined by the Siegel test statistics.}
\label{fig5}
\end{figure}
%
\subsection{Geomagnetic Field Variations at Indian and 210 MM Stations}
\label{S-Indian}
Fig.\ref{deltaH} shows the H variations, from 0700--1700 UT on May 03, 
1998, for seven Indian stations (left panel) and seventeen stations of 
210 MM network (right panel) with the magnetic latitude (ML) of 
each ground station indicated at the right of each curve.  The magnetic 
latitudes of Indian and 210 MM network stations shown in Fig.\ref{deltaH} 
have already been listed in Tables \ref{T-1} and \ref{T-2}. It is evident 
from Fig.\ref{deltaH} that the H variations from 0700--1700 UT for both 
the Indian (left hand panels of Fig.\ref{deltaH}) as well as the 210 MM 
stations (right hand panels of Fig.\ref{deltaH}), had shown increases 
corresponding to the pulses P$_{1}$, P$_{2}$, P$_{3}$, and P$_{4}$.  
Only for the case of pulse P$_{3}$, as already noticed from the variation 
of the AE and AL indices shown in Fig.\ref{fig3}, we found signatures of 
positive and negative bays (e.g., \citep{RGP94b}) on inspection of the H 
variations of 210 MM stations. The H variations at the Indian and the 210 MM 
stations are synchronous at all latitudes, and have been marked  by the 
vertical dashed lines in each panel of Fig.\ref{deltaH}.  It is to be noted 
that these deviations in the H field, as stated earlier, have clearly shown 
a one-to-one correspondence with the episodic enhancements in the solar wind 
density as shown in Fig.\ref{fig3}.
\subsection{Discrete Frequencies of Density and Geomagnetic Field Fluctuations}
\label{S-freq}
It is evident from Fig.\ref{fig3} that, during the period from 0700--1700 UT 
on May 03, 1998, the solar wind density changes and geomagnetic field variations 
were quite similar.  The same also can be clearly seen from their residuals, as 
shown in the top two panels of Fig.\ref{fig5}.  Though the existence of 
quasi-periodic variations is clear from the residuals of density and SYM-H index, 
however, it is difficult to pinpoint the discrete frequencies at which they occurred.  
We, thus, subjected the residuals of solar wind density and geomagnetic field 
variations to a Lomb--Scargle Fourier transform \citep{Lom76,Sca82,Sca89}) to find the 
discrete peaks.  

The bottom two panels of Fig.\ref{fig5} show the Lomb-Scargle normalized periodograms 
of solar wind density and SYM-H field obtained for the period from 0700--1700 UT 
on May 03.  The significant frequency components were determined using the Siegel test 
statistics \citep{Sie80,BiJ14a}.  The black horizontal lines in the bottom two panels 
of Fig.\ref{fig5} are drawn at the critical level determined by the Siegel test.  From 
Fig.\ref{fig5}, it is clear that the density and geomagnetic field pulsations have 
peaks near frequencies, {\it{f}} = 0.48, 0.65, and 0.75 mHz, or periods, 
T = 35, 26, and 22 min.  The frequency peaks reported here belong to the long period or 
ULF oscillations.  In addition, to the common frequencies, the density data also exhibit 
a peak near {\it{f}} = 0.4 mHz, while the SYM-H field show peaks near {\it{f}} = 0.54 
and 0.8 mHz.  As discussed earlier, the periodic density variations are actually 
PDS that are present in the solar wind.  The PDS are formed near the solar 
surface, and convect with the ambient solar wind to 1 AU.  Based on this fact, we computed 
the length-scales of the PDS, using the method followed by \cite{KSp03} and \cite{VKS08}, 
defined as the wavelength, L = $<{V_{x}}>$/{\it{f}}, where $V_{x}$ is the solar wind velocity 
and f is the observed discrete frequency.  The radial scale sizes of the density structures 
were thus found to be occurred at L = 600, 704, 756, 947, and 1160 Mm.
\section{Solar Sources of the Event}
\label{S-sour}
\subsection{Flare and CME Observations}
\label{S-flare}
From the end of April to the beginning of May, 1998, several flares/CMEs erupted 
on the Sun \citep{ThC00}.  A ``{\it{halo}}" CME was captured by Extreme Ultraviolet 
Imaging Telescope (EIT) on board {\it{SOHO}} on April 29 at 1600 UT, and by {\it{Large Angle 
and Spectrometric Coronagraph}} ({\it{LASCO}}; \cite{BrH95}) instrument on board {\it{SOHO}} 
on April 29 at 1658 UT \citep{BaW02,MaS02}.  Just prior to the eruption of the 
CME, a M-class flare (M6.8) erupted at 1606 UT on April 29. It has been reported that 
though the flare and filament eruption occurred in the southeast (S17E16) of the disk center 
from an active region AR 8210, the CME moved across the equator and appeared in the northeast 
limb \citep{WaG00}. The shock driven by the CME on April 29 reached at 1 AU at 2123 UT on May 01, 
while the leading edge of the CME arrived at 0300 UT on May 02 \citep{MaS02}.  A X-class flare and 
another fast halo CME were, respectively, detected in SOHO/EIT at 1340 UT and in LASCO/C2 at 
1406 UT on May 02. The flare and CME were produced in the southwest (S15 W15) from 
AR 8210 \citep{ThC00}, however, the CME moved across the equator and appeared in the 
northwest limb as seen from LASCO/C2 images \citep{WaY02}. The shock driven by the second 
CME traveled along the back of the first CME, and arrived at 1 AU at 1700 UT on May 03.
\section{In-situ Magnetic Field and Remote-sensed IPS Observations}
\label{S-ips}
A study of the observation of bidirectional electron events from the 
Electron, Proton, Alpha Monitor on board {\it{ACE}} ({\it{ACE/EPAM}}) \citep{MaS02} 
has reported that the magnetic field lines at 1 AU during the ICME traversal 
were magnetically anchored to the Sun.  To verify the background IMF-conditions associated 
with the event, we traced back {\it{in-situ}} measurements of solar wind speed (and the 
associated IMF) from the {\it{ACE}} spacecraft at 1 AU to the source surface at 
2.5 R$_{\odot}$, assuming constant velocities and radial expansion along the Parker's spirals.  
The arrival of CMEs, in the present case, could have actually caused deviations in the 
Parker's spiral resulting in the highly non-radial flows \citep{OCa04}.  However, we could 
still use the trace back method considering the fact that the source regions of flows (and 
the IMF) determined would be within reasonable errors.  We verified (though not shown here) 
from the deviations of the solar wind flow velocities estimated using {\it{in-situ}} 
measurements of {\it{ACE}} spacecraft that the flows during May 02--04 were largely radial.  
They only became highly non-radial right after the arrival of the second CME on May 04.  Also, 
the value of $\phi$ remained around 135$^{\circ}$ or 315$^{\circ}$, as shown in the bottommost 
panel of Fig.\ref{fig2}, indicating a steady state IMF configuration.

The heliographic latitude and longitude of the source regions of the IMF at 2.5 R$_{\odot}$ 
were first obtained by transforming the available positions of {\it{ACE}} spacecraft at a 
particular time from the GSE coordinates to Carrington coordinates.  The heliographic 
latitude at the source surface is generally same as that of the heliographic latitude of the 
{\it{ACE}} spacecraft.  However, the heliographic longitude at the source surface was obtained 
by adding an offset to the heliographic longitude of {\it{ACE}} at 1 AU \citep{HuY14}. 
The offset was determined by estimating the longitude through which the Sun had rotated 
during the solar wind propagation time from the source surface to 1 AU using the velocity 
trace back method.

\begin{figure}[ht]
\vspace{9.5cm}
\includegraphics{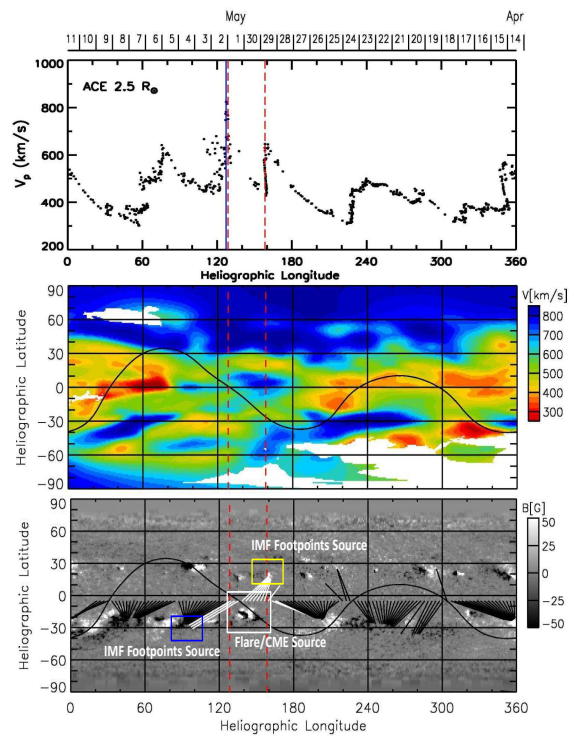}
\caption{The topmost panel shows {\it{ACE}} measurements of the traced back solar wind 
velocities to the source surface at 2.5 R$_{\odot}$ as function of heliographic 
longitudes. The vertically oriented red dashed parallel lines indicate the trace back 
heliographic longitudes of 128$^{\circ}$ and 158$^{\circ}$ corresponding to the prolonged 
period of southward IMF-B$_{z}$, while the vertical blue solid line marks the trace back 
period corresponds to the onset of SI.  Dates of CMP are indicated at the top.  The 
middle panel shows a tomographic synoptic velocity map for CR1935, projected onto the 
source surface at 2.5 R$_{\odot}$, obtained using ISEE IPS data.  The curved solid line 
is the magnetic neutral line.  The lowermost panel shows a synoptic map for CR1935, made 
using magnetograms from the {\it{MDI}} instrument on board {\it{SOHO}}. The heliographic 
longitudes are marked at the bottom of the map.  Regions of large magnetic field 
strength corresponding to active region locations are shown as black and white patches 
to distinguish the two magnetic polarities, while the curved solid line is the magnetic 
neutral line.  Also shown are converging lines in black which join the potential field 
computed source surface magnetic fields with their corresponding photospheric foot points. 
The converging lines in white correspond to the fields of the CMP date from April 29--May 
01. The locations of the footpoints on the photosphere are shown inside the rectangular 
boxes in yellow and blue, while the source region of the flare and CME are shown inside 
a rectangular box in white.}
\label{fig6}
\end{figure}

The topmost panel of Fig.\ref{fig6} shows {\it{ACE}} measurements of proton 
velocities traced back to the source surface at 2.5 R$_{\odot}$ as a function of 
heliographic longitude.  The red dashed vertical lines, marked at the heliographic 
longitudes of 128$^{\circ}$ and 158$^{\circ}$, indicates the day of year (DOY) 
119.72--121.98 (1716 UT, Apr 29--May 01, 2331 UT) corresponding to the trace back locations of 
the solar wind velocities (or the associated IMF) at 1 AU, during our period of interest 
from the DOY 122.39--124.22 (0920 UT, May 02--0520 UT, May 04).  
\protect
\noindent The heliographic latitude of the traced back velocities/IMF at the 
start and the end of the period of interest was found to be 
$\sim$ -4$^{\circ}$ and $\sim$ -5$^{\circ}$, respectively.  
The blue vertical solid line indicates the trace back locations of the IMF 
corresponding to the onset of a SI prior to a strong geomagnetic storm on 
May 04. Dates of Central Meridian Passage (CMP) are indicated at the top of the 
uppermost panel of Fig.\ref{fig6}.  From the topmost panel of Fig.\ref{fig6}, 
it is clear that during this period the proton velocity showed a steady increase 
from 400 km$s^{-1}$ to 820 km$s^{-1}$ without much fluctuations, indicating the 
increase in velocity could be due to the eruption of CME on April 29.

The middle panel of Fig.\ref{fig6} shows a tomographic synoptic velocity map for CR 
1935. The velocity map was obtained using IPS observations from the ISEE, Japan 
and represent solar wind velocities projected on to the source surface at 2.5 R$_{\odot}$.  
The white regions on the map indicate data gaps while the color bar on the right 
shows the velocity of solar wind flow.  The curved black line on the map is the 
source surface magnetic neutral line (MNL).  It is clear from the middle panel of 
Fig.\ref{fig6} that the solar wind flow, in the trace-back region, was largely 
dominated by velocity flows between 500--600 km $s^{-1}$. Also, a high-velocity flow was 
observed north of the equator which is probably an extension of the high velocity 
flows from the high-latitude polar coronal holes.

The lowermost panel of Fig.\ref{fig6} shows a synoptic magnetogram during CR1935, 
from the {\it{MDI}} instrument on board the {\it{SOHO}} spacecraft.  The black 
and white patches represent regions of strong magnetic fields, corresponding to active 
region locations that distinguish the two opposite magnetic polarities. The curved black 
line on the map is the source surface MNL, while the two dashed vertical 
parallel lines bracket the trace back locations.  A large active region is seen to be 
located to the south of the neutral line in the trace back period, which has been identified 
as the source of the flare and CME that erupted on April 29 and May 02, 1998 \citep{ThC00}. 
The converging black lines on the synoptic magnetogram join magnetic fields at the source 
surface to the photosphere. The magnetic fields were computed using a potential field model 
by \cite{HKo99}, and they lie in an equally spaced grid along the equator, while their 
radially back-projected photospheric foot points lie in tightly bunched regions associated 
with active regions north and south of the equator.  These footpoints were the source 
regions of the open magnetic field lines connected from the Sun to 1 AU at the time of 
event. The lines in white mark the field lines corresponding to the period with CMP dates 
from April 29--May 01, 1998.  It is important to note that the converging field lines shown 
in white in the lowermost panel of Fig.\ref{fig6} were found to be associated with active 
regions both north and south of the equator.  The locations of the footpoints on 
the photosphere are shown inside the rectangular boxes in yellow and blue, while the source region 
of the flare and CME are shown inside a rectangular box in white. The magnetic field strength within 
the pixel areas around the first IMF footpoint location has been found to be weaker, while 
the field strength around the second IMF footpoint location has been found to be stronger.
It is evident from the middle panel of Fig.\ref{fig6} that the region of traced 
field lines to the north of the equator was dominated by large median velocity flows of 
approximately 600--800 km$s^{-1}$.
%
\protect
\begin{figure}[ht]
\vspace{7.5cm}
\includegraphics{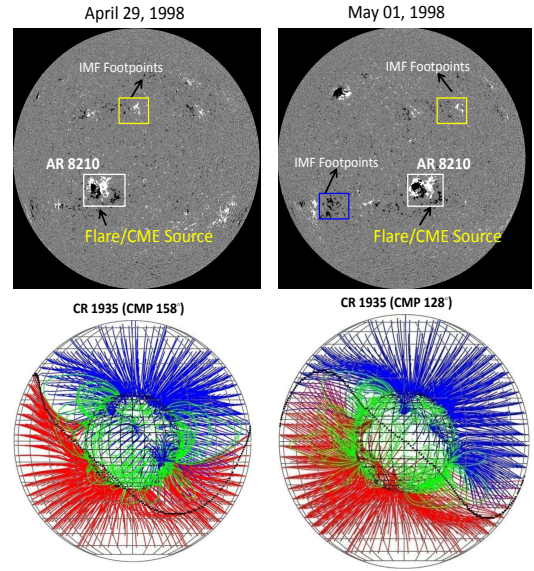}
\caption{Maps of the solar photosphere on April 29 and May 01, 1998 are 
shown in the upper left and right-hand parts, respectively, indicating the 
locations of the source regions. The location of AR8210 is indicated by a 
large rectangular box in both the upper parts, while the IMF footpoint 
locations are indicated by the small rectangular boxes in yellow and 
blue in both the upper parts. The lower two parts show the three-dimensional 
structure of the coronal magnetic fields for CR1935 on April 29, 1998 
(left-hand part) and May 01, 1998 (right-hand part). The fields were 
computed using a potential field model and are viewed from 
158$^{\circ}$ (left) and 128$^{\circ}$ (right) in Carrington 
longitude. The blue and red lines denote the outward and inward polarities, 
respectively, and are shown projected on to the source surface at 2.5 R$_{\odot}$. 
The green lines denote the close field lines.  Only fields between 5 G and 
250 G on the photosphere are plotted in order not to clutter the figure.  
The open field lines at 2.5 R$_{\odot}$ are plotted in steps of 5$^{\circ}$ 
in longitude and latitude.}
\label{fig7}
\end{figure}	      
%
\subsection{Active Region Locations} 
\label{S-location} 

The upper two panels of Fig.\ref{fig7} show the location of active regions 
on the sun on April 29, 1998 (upper left) and May 01, 1998 (upper right).  The 
active region AR 8210, was the source region of the flare/CME that erupted on
April 29, is shown inside a large rectangular box in white in the top 
left panel. The active region is located close to the central meridian at a 
longitude of 16$^{\circ}$E to the south of the equator on April 29.  The 
footpoint of the IMF, at the start of the traced back interval, on April 29 is 
associated with a small active region shown inside a small rectangular box 
in yellow in the top left panel. This active region is located in 
close proximity to the central meridian on April 29. The top right panel of 
Fig.\ref{fig7} shows the positions of the active regions on May 01. The 
location of the IMF footpoint, at the end of the traced back interval, on May 
01 is associated with the active region on the south of the equator, shown in a 
small rectangular box in blue. 

The lower two panels of Fig.\ref{fig7} show the three-dimensional structure of the 
computed coronal magnetic fields on April 29, 1998 (lower left panel) and May 
01, 1998 (lower right panel) respectively, as viewed from a Carrington longitude of 
158$^{\circ}$ (left) and 128$^{\circ}$ (right) corresponding to the start and 
the end of the traced back interval. The coronal magnetic fields can't be measured 
directly, and so were computed using photospheric magnetic fields as inputs to 
potential field extrapolations \citep{HKo99}. The magnetic field lines from the 
potential field computations fan out from tightly bunched regions associated with 
active regions north and south of the equator with the blue and red lines denoting 
the outward (positive) and inward (negative) open fields, respectively.  The green 
lines denote closed field lines, while the solid black line is the MNL.  Only fields 
between 5 G and 250 G on the photosphere have been plotted 
in order not to clutter the figure, and the open field lines at 2.5 R$_{\odot}$ are 
plotted in steps of 5$^{\circ}$ in longitude and latitude.  From Fig.\ref{fig7} 
(lower panels) it is evident that apart from the polar coronal hole regions 
showing the open field lines, the other open field lines were associated with the 
small active region north to the equator, which is the source region of the 
IMF footpoints of the prolonged southward IMF-B$_{z}$ interval, as shown inside the small 
rectangular box in yellow in the top left panel.  It is important to note that the open 
field lines associated with the region were found to be positive and remained constantly 
north of the MNL both on April 29 and May 01. This implies that the large-scale open 
field structures associated with the region didn't change their polarity and remained 
radial beyond the source surface for most of the time during the trace back period 
corresponding to the prolonged southward IMF event. 

It can be seen from Fig.\ref{fig7} (lower right panel) that the MNL shifted 
towards the north-west on May 01.  The open magnetic field structures directed inward 
were found to the south of the MNL associated with a bunch of closed magnetic field lines 
and loops.  As mentioned earlier, they were the footpoint locations of the traced back 
IMF at the end of the prolonged southward IMF interval, shown inside a small rectangular 
box in blue in the top right part, associated with an active region on the southeast 
of the solar disk center.
\section{Discussion}
\label{S-discuss}
Our analysis of the north-south component of interplanetary magnetic field at 
1 AU during geomagnetic storm conditions in the period May 02--04, 1998 showed 
a nearly 44 hour long or unusually prolonged weakly southward IMF-B$_{z}$ in 
the time interval from 0920 UT, May 02, 1998 to 0520 UT, May 04, 1998. This is 
for the first time that such an unusually long duration southward IMF-B$_{z}$
event has been reported.  

\subsection{Interplanetary Sources}
\label{S-IP}
From {\it{in-situ}} measurements, at 1 AU, of solar wind plasma and magnetic 
field, it is evident that the prolonged IMF-B$_{z}$ event was associated with 
an ICME identified from May 02, 0500 UT--May 04, 0100 UT.  The ICME was actually 
the counterpart of a large ``{\it{halo}}" CME that erupted on April 29, 1998 at 
1658 UT \citep{ThC00,MaS02}.  Prior to the CME eruption, a M6.8 flare erupted at 
1606 UT.  The shock driven by the flare and CME led to the forward shock at 2123 UT 
on May 01 and the reverse shock at 0300 UT on May 02, respectively.  
The arrival of the forward shock caused the compression of 
the upstream plasma and led to the fluctuations in the IMF components.  Subsequently, 
the draping of the field lines around the CME in the shocked plasma produced the 
northward IMF-B$_{z}$ fields.  The arrival of the ICME, marked by a reverse shock, 
resulted in an out-of-ecliptic IMF component, that led to the first southward turning 
of the IMF-B$_{z}$.  In a study of the outward propagation of the flare-associated CME, 
\cite{HZh92} reported the preferential existence of the southward IMF-B$_{z}$ 
component at 1 AU.  It is, thus, the arrival of the ICME at 1 AU, driven by 
the flare-associated CME of April 29, that led to the first southward 
turning of the IMF-B$_{z}$ observed at 0920 UT on May 02.

The ICME interval at 1 AU contained a well-defined MC observed from 1300 UT, May 02 
to 1200 UT, May 03, for over 24 hours.  Given such long hours of traversal of MC, 
we estimated the radial width of the MC and found it to be nearly 0.28 AU.  
Also, {\it{in-situ}} observation of the monotonic 
decreases in the solar wind speed and the magnitude of the magnetic field shows that 
the MC was expanding in the IP space.  As mentioned earlier, it has been reported by 
\cite{MaS02} that the magnetic field lines during the ICME passage were anchored to 
the Sun.  Also, as discussed above, the fields were already southward, prior to the 
MC, due to the arrival of the ICME.  Thus, as the MC approached, the southward 
field lines draped around the MC and led them to be further southward and intense 
\citep{GoJ94}.  Since the source region of the CME that erupted on April 29 was in the south 
of the disk center, the internal field of the CME/MC should have a southward directed 
field \citep{McG89}.  It is known that most MCs while propagating 
through the IP space don't change their polarity \citep{Mar86}, so the southward 
directed internal field could result in a southward IMF component at 1 AU. Earlier 
observations \citep{ZhH98,ZMo14} have also reported that the most southward IMF-B$_{z}$ 
events are associated with ICME having MC signatures.  It is, thus, expected that the 
prolonged weakly southward magnetic fields, for nearly 24 hours, during May 03, was a 
result of the expanding large MC associated with the ICME passage. Also, the long 
duration of MC can be studied by comparing the magnetic flux geometry of the MC at the 
location of the encounter with the spacecraft, obtained using the magnetic flux rope models 
\citep{Bur88,Mar00,MLe07,MaA15}, to the observed magnetic field variations at 1 AU.

The arrival of a second forward shock and an IP discontinuity at 1 AU 
\citep{SkB99}, driven by a second flare-associated CME erupted on May 02, 1406 UT 
\citep{SkB99,MaS02}, compressed the pre-existing southward IMF-B$_{z}$ component 
\citep{TsS88,GoJ94}. This, in turn, resulted in a strong southward IMF-B$_{z}$ as 
noticed on May 04, 0200 UT.  

\subsection{Solar Sources}
\label{S-solar}
Two flare-associated CMEs erupted, respectively, on April 29 and May 02.  The source region 
of both of the CMEs was a large active region AR 8210 \citep{ThC00}.  We tracked source 
regions of the solar wind flows associated with the event, using the trace back technique 
of constant velocities along Archimedean spirals, back to the solar source surface at 
2.5 R$_{\odot}$. The source locations of solar wind flows, corresponding to the start and 
the end of the IMF interval, were identified to the south of the equator and between the heliographic 
longitudes of 128$^{\circ}$ and 158$^{\circ}$.  From observations of solar 
wind velocity from {\it{ISEE}}/IPS observatory and the photospheric magnetic fields from 
{\it{SOHO/MDI}}, we were able to show that the traced back flows were associated 
with an active region AR 8210 at a longitude of 16$^{\circ}$S.  It is to be noted that 
the active region is the source region of the flares/CMEs erupted on April 29 and May 02.  
A positive SI was observed at the end of the present long duration southward IMF event. 
This was caused due to a shock driven by a second faster CME, which erupted on May 
02, and arrived at the back of the first CME just prior to the end of the first ICME 
interval.  The arrival of the second shock just before the end of the first ICME 
raised the question whether the eruption of the second CME had destroyed the magnetic 
configuration of the field lines during the trace back period, and whether 
information of the background IMF conditions on the solar surface was destroyed. 
From our analysis, however, we showed that the eruption of CME took place right after our 
trace back period indicating that the source regions of the IMF determined by the 
velocity trace back technique would be valid \citep{KoT13}.  It is also to be noted 
that the active region AR 8210 remained south of the magnetic neutral line (MNL) during 
the whole trace back period indicating no change in the IMF configuration.

Using the field-line tracing of the computed coronal magnetic fields from the 
solar surface to the photosphere, we were able to locate the footpoints of the 
IMF source region at the start of the prolonged southward IMF-B$_{z}$ event at 1 AU to a 
small unnamed active region on April 29, lying close to the central meridian and to the 
north of the equator.  The footpoints of the IMF source regions at the end of 
the prolonged southward IMF-B$_{z}$ event at 1 AU were found to be associated with a 
region in the close-proximity of an active region, located to the south of the 
equator.  The magnetic field lines during the passage of the ICME were anchored to the 
Sun, thus, the footpoints of the traced back IMF were actually the 
foot regions of the ICME observed at 1 AU.  Further, the potential field computations 
of source surface magnetic fields on April 29 and May 01 showed open magnetic field 
structures with positive polarity emanating from this unnamed active region.  The open 
field lines, which remained constantly to the north of the heliospheric current sheet both 
on April 29 and May 01, clearly suggest no change in their polarity during the trace back 
period, from April 29--May 01.  It is also known from observations that, depending on the 
hemisphere, the open magnetic field lines spiral either inward or outward, and the 
configuration changes, generally, either during periodic solar polar reversals occurring 
at the maximum of each solar cycle \citep{JaF05,JaF08,JDM08} or due to the wake of CME 
eruption \citep{KoT13}.  However, the polarity reversal of solar cycle 23 occurred 
in CR 1949 while our event was during CR1935.  Also, the second CME was erupted on May 
02, after our trace back period and after the first CME which erupted on April 29.  
From our observations, it is evident that the sudden change in proton velocity (due to 
the eruption of the CME on April 29) was seen before or around the same period when the 
IMF-B$_{z}$ at 1 AU became southward. Therefore, no change in magnetic field configuration 
can be expected.  It is noticed however, from our field-line tracing that the IMF 
showed a change, towards the end of the southward IMF-B$_{z}$ interval, in their footpoint 
locations from the north to the south.  Also, the potential field computations of source 
surface magnetic fields on May 01 showed open magnetic field lines with negative polarity 
emanating close to the southeast active region.  This is consistent with observation of the 
IMF crossing at 1 AU (change in the value of $\phi$ from positive to negative as shown 
in Fig.\ref{fig2}) during the ICME passage.
\protect
\begin{figure*}[ht]
\vspace{6.5cm}
\includegraphics{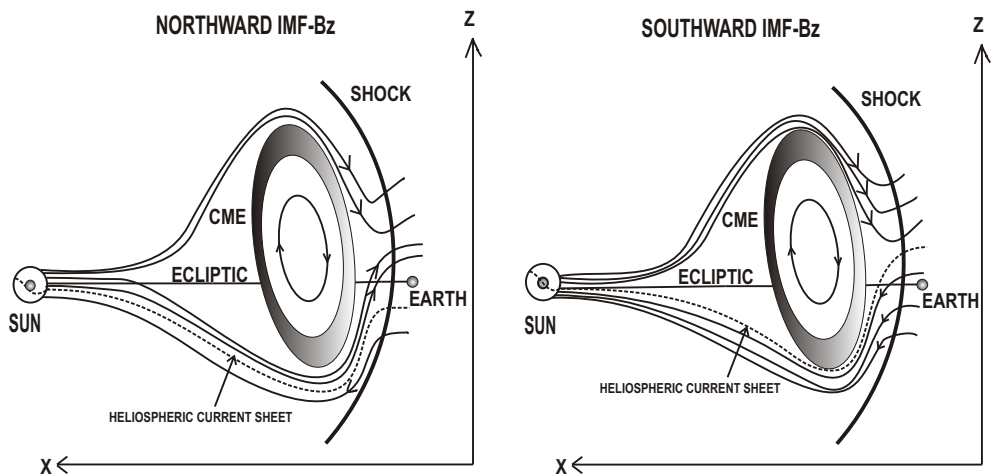}
\caption{A schematic showing the draping of the field lines around the CME 
in the IP space. (Left panel) The CME of April 29 propagating in the north 
of the ecliptic led to the draping of the field lines in the shocked 
plasma, which were directed outward, and generated the northward IMF-B$_{z}$ 
fields at 1 AU. (Right panel) The CME of May 02 propagating in the north 
of the ecliptic led to the draping of the field lines in the shocked plasma, 
which were directed inward (sunward) and generated the southward IMF-B$_{z}$ 
fields at 1 AU. Both the CME had southward directed internal fields as 
their source regions were located to the south of the solar disk. The position 
of the HCS on the Sun and in the IP space, have been shown by a dotted line.  
The HCS separates the positive and negative IMF field lines.}
\label{fig8}
\end{figure*}      
%

Fig.\ref{fig8} shows a schematic of the draping of the field lines around the 
CME plasma.  As mentioned earlier, the first CME, erupted on April 29, was 
propagating northward \citep{WaG00}, so the center of the CME would be to the north 
of the ecliptic.  Also, the IMF field lines were directed away from 
the Sun (outward), while the HCS was to the south of the equator.  Such a condition 
with the CME propagating to the north of the ecliptic and with the field lines 
radially outward has been shown in the left panel of Fig.\ref{fig8}.  It can be seen 
from Fig.\ref{fig8} that the draping of the field lines around the CME, in the shocked 
plasma, has resulted in northward IMF-B$_{z}$ fields \citep{GMc87,McG89}.  This is 
consistent with the observed northward IMF-B$_{z}$ fields noticed after the post-shock 
early on May 02.  The CME internal field, which was southward directed, resulted in the 
southward IMF-B$_{z}$ fields at 1 AU.  The second CME, erupted on May 02, was also 
propagating to the north \citep{WaY02}, so the center of the CME would be to the north 
of the ecliptic.  The IMF field lines, as mentioned earlier, on the north were still radially 
outward, while the IMF field lines on the south were radially inward.  As mentioned 
earlier, the HCS moved to the northwest, but was still to the south of the 
equator.  The right panel of Fig.\ref{fig8} presents such a condition, where the CME was 
propagating to the north of the ecliptic and the field lines were radially inward, which 
resulted in southward IMF-B$_{z}$ fields due to the draping of the sunward field 
lines around the CME/MC.  The southward fields when approached the MC within the CME 
having southward directed internal fields would be intensified further.  In fact, 
strong southward IMF-B$_{z}$ fields were observed towards the end of the interval.  Assuming 
the CME moved radially outward from the Sun, the IMF fields originally 
anchored to the Sun southward (northward) of the CME source location would still be traced 
back to the Sun southward (northward) of the CME \citep{McG89}.  It is to be noted that the 
CMEs of April 29 and May 02, erupted from the same active region, AR 8210, which was located 
to the south of the disk center.  Thus, when we tracked down the radial IMF fields at the 
start and the end of the interval, the IMF footpoints were located to the north and the south 
of the equator, respectively, as expected.

It is, thus, clear that the passage of the ICMEs in the interplanetary medium played a 
crucial role that led to the unusually prolonged southward IMF-B$_{z}$ condition.  
The open field lines from the northern active region, which were anchored to the Sun and 
remained constantly radially outward during the whole interval, explained the observed 
polarity changes of the IMF-B$_{z}$ fields within the shocked plasma. 
\subsection{Geomagnetic Activity}
\label{S-pulse}
The IMF-B$_{z}$ remaining southward for prolonged hours during May 02--04 
led to a moderate storm and a strong storm, respectively, early on May 02 and 
early on May 04. In between the storms during the weakly prolonged quasi-stable 
IMF-B$_{z}$ condition (B$_{z}$ $\geq$ -5nT) on May 03, we also found multiple peaks in 
the SYM-H with corresponding enhancements in the solar wind dynamic pressure 
(and solar wind density) between 0700 UT and 1700 UT on May 03, 1998.  These 
pulses are unique in a sense that both the IMF-B$_{z}$ and solar wind velocity 
barely showed any variation.  Thus, any significant change in the ground level 
magnetic fields could be attributed to the solar wind pressure variations 
(or density variations). The multiple peaks identified both at low-latitude 
Indian stations and 210 MM network of stations during the above period provided 
a clear evidence that these enhancements were global in nature.  Since perturbations 
in SYM-H index, usually represent mean variations in the symmetric ring current 
(the westward symmetric electric current that flows along the geomagnetic equator), 
it is clear that the prolonged weakly southward IMF-B$_{z}$ preconditions on May 
02-04, 1998 developed a quasi-stable ring current, promoted by solar wind-magnetosphere 
coupling for a long period.  

In the presence of the strong southward IMF-B$_{z}$ 
for long intervals, the solar wind flux has been continuously loaded to the 
magnetosphere/magnetotail causing a magnetospheric convection on the 
nightside. This, in turn, leads to transport and acceleration of particles 
from the magnetotail and the plasmasheet towards the Earth's inner magnetosphere 
\citep{KaB98,DaT99} forming the storm-time ring current.  In the present study, 
early on May 02, when the southward IMF-B$_{z}$ was relatively stronger, the 
ring current was intensified, as seen by the relative depression in SYM-H index, 
due to the enhanced convection.  Early on May 03 while the southward IMF-B$_{z}$ 
was relatively weaker, it is not expected to form a ring current.  Under 
these weak IMF-B$_{z}$ conditions, the density variations in the solar wind 
can be a driver for the ring current \citep{CLS94,JoF98,BoT98,SmT99,WCL03}.  
The plasma sheet density in the near-Earth plasma sheet regions, which is 
well correlated to the density in the solar wind \citep{TeF97,BTM97,BoT98}, 
supplies the energy and form the ring current \citep{ThB98,KoJ98,LiK99}. 
In the present case, the increases in the solar wind density, from 10--60 cm$^{-3}$ 
early on May 03, when the IMF-B$_{z}$ was weak but still southward, perhaps 
led to increased number of particles entering the magnetosphere.  This, in turn, 
increased the energy accumulated in the magnetotail and plasmasheet, even when 
the energy injection began to decline due to the saturated magnetospheric convection.  
Subsequently, the solar wind dynamic pressure enhancements, corresponding to density 
variations during 0700--1700 UT on May 03, formed a quasi-stable ring current.  
We believe that the multiple peaks in the SYM-H index were manifestations of 
the multiple peaks developed in the quasi-stable ring current.  Our study, thus, 
indicates that P$_{dyn}$ variations owing to density changes might play an 
important role in the development of ring current during the quasi-steady 
prolonged southward IMF-B$_{z}$ condition.  In a similar study, \cite{ArF93} 
showed the effect of the solar wind dynamic pressure, caused by solar wind 
density variations, on the geomagnetic field through study of changes in the 
Dst index. The study \citep{TaS05} of the solar wind pre-conditions 
during the different magnetotail phenomena corresponding to the prolonged 
southward IMF-B$_{z}$ intervals reported that the magnetospheric 
convection is more likely when the mean southward IMF-B$_{z}$ is $\geq$ -5 nT, 
while loading and unloading is more likely when IMF-B$_{z}$ is $\leq$ -5 nT.  
In the present study, we found that the southward IMF-B$_{z}$ was $\geq$ -5 nT 
during the weakly southward IMF-B$_{z}$ condition on May 03.  It indicates an 
occurrence of a night side steady magnetospheric convection during the period 
which, thus, led to the development of a quasi-stable ring current owing to the 
density variations.

Also, our periodogram analysis suggested the presence of the discrete frequencies 
in the solar wind density variations, during 0700--1700 UT, May 03, near {\it{f}} 
= 0.48, 0.65, and 0.75 mHz, or T = 35, 26, and 22 min.  It is known from earlier 
observations (\cite{KSS02,KSp03,VKS09}) that such long period and sub-mHz ULF 
frequencies oscillations have been existed in magnetospheric fields.  Confirming 
the fact, we also found the similar discrete frequencies in the ground magnetic 
fields or the symmetric ring current strength (SYM-H fields).  The frequencies 
obtained here in solar wind density variations are in agreement with those reported 
by \cite{VKS09}.  Also, the estimation of their radial length scales showed that 
the associated PDS of the density variations have significant power at wavelengths, 
L = 600, 704, 756, 947, and 1160 Mm.  The observed length scales of the 
PDS estimated here are larger in sizes than those reported by \cite{VKS08}.  Our 
spectral analysis results fit well into the global picture that shows that the ULF 
oscillations of the solar wind density variations act as the drivers of the 
magnetospheric field oscillations.  The density variations could be attributed 
to the significantly increased values of the He$^{++}$/H$^{+}$ ratio \citep{SkB99} 
observed at 1 AU, between 1800 UT on May 02 and 1700 UT on May 03, though the origin 
of the density pulses is need to be investigated. The origin of the associated PDS 
with the density variations has been well studied \citep{VSK09,ViS10,VVo15}, which 
revealed that these can be due to variations in the solar coronal plasma, which 
convect with the ambient solar wind from the Sun to 1 AU.  The increased 
He$^{++}$/H$^{+}$ ratio at 1 AU, during the prolonged southward IMF-B$_{z}$ 
interval, indicates that the PDS could be of the solar origin, 
which convected with the solar wind during the outward propagation of the CME from the Sun 
to 1 AU \citep{VSK09}. The details of those, however, requires a separate further 
study and will be investigated in a future work.

It is generally understood that the Earth's magnetotail during such long hours of the 
southward IMF-B$_{z}$ intervals respond in different manners such as isolated loading 
and unloading, periodic loading and unloading, and steady magnetospheric convection 
(\cite{TaS05} and references therein).  A series of substorms, known as periodic 
substorms, can occur when the southward IMF-B$_{z}$ is southward for several hours 
\citep{Hua02,HuR03,HuF03}. Since the magnetosphere is open for long hours during these 
IMF-B$_{z}$ conditions, solar wind variations can make their way into the magnetosphere, 
and subsequently to the ground causing disturbances at low and mid-latitudes with periods 
of nearly 2--3 hours.  These substorms are usually associated with periodic loading 
and unloading of plasma particle fluxes and are commonly referred to as saw tooth events 
\citep{Hen04}.  They, in turn, can contribute to disturbances recorded at ground magnetic 
stations.  Thus, one of the possible reasons for the occurrence of the multiple pulses on 
May 03, 1998 may be due to the substorm activity.  So it is important to delineate the 
multiple processes occurring together.  We have taken care to verify the energetic proton 
fluxes from all the available Los Alamos National Laboratory ({\it{LANL}}) spacecraft to 
see if there were any periodic injections of fluxes.  However, except for the pulse P$_{3}$, 
we didn't find such flux injections corresponding to the onset of pulses P$_{1}$, P$_{2}$, 
and P$_{4}$, rejecting the possibility of any periodic substorms during that period. 

Besides the precursor southward IMF of more than 10 hours, typical signatures of 
magnetospheric/magnetotail preconditioning were found prior to the onset of the storm on 
May 04 such as a SI, constant values of SYM-H with $\leq$ -50 nT, and a fairly constant 
substorm activity with the values of AE between 500 and 1000 nT.  The great storms 
reported by \citep{TsL92} also reported similar magnetospheric/magnetotail 
pre-conditioning with Dst $\leq$ -50 nT and AE $\geq$ 500 nT prior to the onset of the 
great magnetic storm.  However, a careful inspection of the AE and AL indices, 
in the present study, with AE $\geq$ -2500 nT showed that they were not supersubstorms as 
discussed in \cite{TsH15}.
\section{Summary and Conclusion}
\label{S-cons}
The present work is important as it is for the first time such a prolonged weakly 
southward IMF-B$_{z}$ of nearly 44 hours has been reported and the cause of this 
has been addressed in detail by investigating both the interplanetary and 
solar sources.  The interplanetary plasma and magnetic field measurements 
at 1 AU showed the passage of two ICMEs during the event.  The shocks and the 
magnetic cloud driven by the first ICME led to the observed southward IMF-B$_{z}$ 
during May 02--04.  The sharp southward IMF-B$_{z}$ near the end of the event was 
due to the second ICME and its upstream shock.  Investigation of the solar sources 
indicated that the first ICME was due to a fast halo CME which erupted on April 29 and 
the second ICME was due to a fast halo CME erupted on May 02. It is to be noted 
that both the CMEs erupted from the same active region AR 8210.

Also, we were able to locate the source region of the solar wind flows and 
the associated IMF during the event to the solar surface at 2.5 R$_{\odot}$, 
located in close proximity of the active region AR 8210.  The field-line 
tracing of the IMF, from the source surface to the photosphere, showed that 
the footpoints of the IMF during the event were associated with 
a small active region on the north and an active region on the south.
Examination of their corresponding magnetic configuration showed open magnetic field 
structures emanating from the northern active region.  These open field lines 
were directed away from the Sun and didn't change their direction during 
the whole trace back period of April 29--May 01, corresponding to the 
event at 1 AU.  The draping of these constantly radial outward field lines,
during the prolonged southward IMF-B$_{z}$ interval, around the CME/MC 
at 1 AU, explained the observed polarity changes of the IMF-B$_{z}$ within the 
shocked plasma.

Additionally, we also investigated multiple global disturbances of the geomagnetic 
fields from 0700--1700 UT on May 03 to understand the underlying cause.  It was 
shown that the multiple density enhancements in the solar wind, within a period of 
only 10 hours, could develop quasi-stable ring currents.  This, in turn, could lead 
to the multiple surface magnetic field deviations, without the contribution from 
either the solar wind velocity or the IMF-B$_{z}$.  Further, the spectral 
analysis of solar wind density and geomagnetic field variations showed 
the presence of the common discrete frequencies at 0.48, 0.65, and 0.75 mHz confirming 
the solar wind density variations as the driver of the global geomagnetic disturbances.  
The work, thus, presents an observational link of solar surface phenomena with that 
of worldwide magnetic response during an unusually prolonged southward IMF-B$_{z}$ 
condition.
%

%
%
%
%
%
%
%

\begin{acknowledgments}
The authors thank King, J.H. and Papatashvilli, N. of Adnet Systems, 
NASA, GSFC, the CDAWeb team, the {\it{ACE}} instrument team, the 
{\it{ACE}} Science Center, the {\it{MDI}} consortia for making data 
available in the public domain via the World Wide Web. {\it{SOHO}} 
is a project of international collaboration between ESA and NASA. 
IPS observations were carried out under the solar wind program of 
ISEE, Nagoya University, Japan.  For IMF and solar wind data used 
can be obtained from \url{http://cdaweb.gsfc.nasa.gov/cgi-bin/eval2.cgi}.  
{\it{ACE}} and {\it{MDI/SOHO}} data are available in the 
public domain (\url{http://www.srl.caltech.edu/ACE/ASC/level2/index.html} 
and \url{http://sohodata.nascom.nasa.gov/cgi-bin/data_query}). Other 
data are available from the authors upon request such as for IPS 
solar wind velocity (fujiki@stelab.nagoya-u.ac.jp), Indian 
magnetograms used in the present paper (susanta@nao.cas.cn) and 
210 MM magnetograms (yoshi@geo.kyushu-u.ac.jp). SKB acknowledges the 
support by the Chinese Academy of Sciences International Talent Scheme, 
Project Number: 2015PM066 (Funded by CHINESE ACADEMY OF SCIENCES 
PRESIDENT'S INTERNATIONAL FELLOWSHIP INITIATIVE). Yihua Yan is 
supported by NSFC Grant: 11433006. The work by A.Y.  was supported 
by MEXT/JSPS KAKENHI Grant Number 15H05815.  One of the authors, JP, 
would like to thank Prof. U.R. Rao for useful suggestions that have 
improved the paper. The authors also thank the reviewers for their 
constructive comments and suggestions that have significantly 
improved the paper.
\end{acknowledgments}

\bibliographystyle{agufull08}
%

%
%
\end{article}

\end{document}